\documentclass[nofootinbib,showpacs,preprintnumbers,
amsmath,amssymb,superscriptaddress]{revtex4}

\usepackage{axodraw}
\usepackage{graphicx}

\begin{document}

%\begin{center}

\title{ On the forward-backward charge asymmetry in $e^+e^-$ -annihilation
into hadrons at high energies}

\vspace*{0.3 cm}

\author{B.I.~Ermolaev}

\altaffiliation[Permanent Address: ]{Ioffe Physico-Technical Institute, 194021
 St.Petersburg, Russia}
\affiliation{CFTC, University of Lisbon
Av. Prof. Gama Pinto 2, P-1649-003 Lisbon, Portugal}
\author{M.~Greco}
\affiliation{Dipartamento di Fisica and INFN, University of Rome III, Italy}
\author{S.I.~Troyan}
\affiliation{St.Petersburg Institute of Nuclear Physics, 
188300 Gatchina, Russia}

%\end{center}

\begin{abstract}

The forward-backward asymmetry in $e^+ e^-$ annihilation into a
quark-antiquark pair is considered in the double-logarithmic approximation
at energies much higher than the masses of the weak bosons.
It is shown that after
 accounting to all orders for the exchange of virtual photons and $W, Z$
-bosons one is
lead to the following effect (asymmetry): quarks with positive electric
charge (e.g. $u$, $\bar{d}$) tend to move in the $e^+$ - direction whereas
quarks with negative charges  (e.g. $d$, $\bar{u}$) tend to
move in the  $e ^-$ - direction. The value of the asymmetry grows with
increasing  energy when the produced quarks are
within a cone with opening angle, in the cmf,
$\theta_0\sim 2M_Z/\sqrt{s}$ around the $e^+e^-$ -beam. Outside this
cone, at $\theta_0 \ll \theta \ll 1$, the
asymmetry is inversely proportional to $\theta$ .
\end{abstract}


\maketitle 

\section{Introduction}

The standard theoretical description of
$e^+e^-$ annihilation into
hadrons at high energies
starts with the sub-process of
$e^+e^-$ -annihilation
into quarks and gluons, which is then
 studied with
perturbative methods. It is usually considered
as mediated by the exchange of all electroweak (EW) bosons:
$e^+e^- \rightarrow \gamma^{*}, Z, W \rightarrow q \bar{q} +$gluons.
One of the most well-known and successful predictions of the Standard
Model is the
forward-backward asymmetry,
which has been studied for many years both theoretically and
experimentally, particularly around the Z boson \cite{1}.
Next linear colliders will explore $e^+e^-$ annihilation at very high
energies, probing further the Standard Model and eventually looking for
New Physics. As it is well known,
pure QED also gives rise to a
 forward-backward
asymmetry even
at low energies, albeit small, due to interference of one-photon
and two-photon exchange diagrams. This effect persists at asymptotically
high energies due to
the  multiphoton contributions  in
higher orders in $\alpha$.
Such  multiphoton contribution in $e^+e^- \rightarrow
 \mu^+\mu^-$  was studied in Refs.~\cite{gg}
in the double-logarithmic (DL) approximation (DLA).
The annihilation process in \cite{gg}
was considered in the following two kinematic regions:

\begin{itemize}
\item[(i)] Forward kinematics, when,  in the center of mass frame (cmf),
the outgoing $\mu^+$
( $\mu^-$ )
goes in the direction of the initial $ e^+ (e^-)$.
\item[(ii)] Backward kinematics, when
the outgoing $\mu^+$  $(\mu^-)$
goes in the  $e^-$ ($e^+$) -direction.
\end{itemize}

These kinematical configurations refer to the case when
 the initial positive (or negative) electrical charges  do not
change  the direction  after the scattering, or they are affected by a
major - almost backward - deviation.
It was shown in \cite{gg} (see also the review \cite{g}) that at high
energies the radiative DL corrections to
the Born amplitudes
are quite different for the forward and the backward kinematics.
As a result, the cross section  of the forward
annihilation dominates over the backward one and therefore there is a
charge forward-backward asymmetry:
positive muons tend to go in the $e^+$ -beam  direction and negative
muons rather follow the direction of $e^-$.

In the present paper we
generalize these results, accounting  also for DL contributions of
multiple W,Z exchanges.   We calculate below the forward and backward
scattering amplitudes for  $e^+e^-$ -annihilation into $q \bar{q}$
-pair at energies much higher than $M_Z$ and estimate the dependence of
the asymmetry on the scattering angle and on the total energy of the
process.  We account for multiphoton
exchanges as well as for multi- $W,Z$ exchanges to all orders in the EW
couplings in DLA.  To this aim we construct and solve in DLA an
infrared evolution equations (IREE) for the backward and forward
scattering amplitudes.  In doing so we follow the approach of
Refs~\cite{kl,el}.  Recently this approach was used\cite{flmm} for
studying the double logarithmic asymptotics for
lepton-antilepton backward scattering, for scattering angles $\theta \sim
\pi$ in the EW theory.
Furthermore, in
 Ref.~\cite{flmm} it is considered the  case where the initial (final)
state consists of a left (right) lepton and its antiparticle.  However,
because there is an essential difference in the EW theory in
the description of the processes
involving the left and the right fermions,  in the
present paper we generalize those results and consider the forward and
backward annihilation
with  both left and right initial and final states.
We also study these
processes in a wider  angular region $\theta \ll 1$. Although we use the
Feynman gauge through this paper, all our results are gauge invariant.
The paper is organized as follows. Instead of calculating
amplitudes of $e^+e^- \to$ quarks directly, we find more convenient
to operate with $SU(2)$ - invariant amplitudes of a more general process,
 the lepton-antilepton annihilation into quark-antiquark pair. Then in
Sect.~2 we introduce such invariant amplitudes and show their relation
to the forward and backward amplitudes for  $e^+e^-$ -annihilation.
The IREE for the invariant amplitudes are constructed in Sect.~3 and
solved in Sect.~4. The solutions are obtained in terms of the Mellin
amplitudes corresponding to collinear kinematics.  The IREE for the
Mellin amplitudes are obtained and solved in Sect.~5.  Then, we define
and estimate the forward-backward asymmetry in Sect.~6.  Finally, we
discuss our results in Sect.~7.

\section{ Invariant amplitudes for lepton-antilepton
annihilation into $q \bar{q}$.}\label{sect1}

We are going to account for the DL effects of
exchanging the EW bosons to $e^+e^-$ -annihilation into
quark-antiquark pairs of different flavours.
When multiple $W$ - exchange is taken into account,
the flavour of the virtual intermediate fermion state is not fixed,
though the
initial and final states of the annihilation are well-defined.
Because the EW theory  organizes all
fermions into doublets of the left particles
and right singlets, this suggests that is more convenient  to
calculate first the
scattering amplitude of a more general process, the
annihilation of a lepton and its antiparticle
into a quark - antiquark pair, and only after  to
specify the flavour of the initial and the final states.
This  turns to be easier because the
effects of the violation of the
initial $SU(2)U(1)$ symmetry are in many respects neglected within the
double-logarithmic accuracy.
On the other hand that also means that the
DLA can be applied safely  only
when the energy of the annihilation is
much higher than $M_Z, M_W$. At such energies  the propagators
of the $SU(2)$ - gauge bosons $W_a$ ($a = 1,2,3$),
$D_{ab}(k) \sim \delta_{ab}/k^2$.
The propagator  of the $U(1)$
-gauge boson $B$ is $1/k^2$ in the same approximation.
The $SU(2)$ vertices of the $W_a$ interaction with
the left fermions are $g\,t^a$, where
$t^a$ are the $SU(2)$ generators and $g$ is the coupling,
whereas the vertex of the interaction of the field $B$
with the left and the right fermions is $g'\,Y/2$,  $Y$ being the
hypercharge and $g'$ being
the coupling. As in the most general process both the initial
and final
particles can be left and/or right, we consider all these cases
separately.

In this Sect. we consider the general case of the
annihilation of the left lepton $l^k(p_1)$
belonging to the doublet $(\nu,~e)$ and  the
antilepton $\bar{l}_i(p_2)$ from the charge conjugated doublet
into the left quark $q^{k'}(p'_1)$ belonging to the doublet $(u, ~d)$
and the antiquark $\bar{q}_{i'}(p'_2)$ from the charge conjugated doublet.
Therefore, the scattering amplitude $M$ of this process is

\begin{equation}
\label{a}
M = \bar{l}_i(-p_2) \, \bar{q}_{i'}(p'_1) \,
\widetilde{M}^{ii'}_{kk'} \, l^k(p_1) \, q^{k'}(-p'_2)
\end{equation}

where   the matrix amplitude $\widetilde{M}_{ii'}^{kk'}$
has to be calculated perturbatevely.

We consider the kinematics where, in the cmf,
both particles of
the produced  pair move close to the lepton-antilepton beam.
It corresponds to two kinematics:

\begin{itemize}
\item[(i)] forward kinematics when
\begin{equation}
\label{smallt}
-t = -(p'_1-p_1)^2 << s = (p_1+p_2)^2 \approx -u = -(p'_2-p_1)^2 ~,
\end{equation}
\item[(ii)] backward kinematics when
\begin{equation}
\label{smallu}
-u = -(p'_2-p_1)^2 << s = (p_1+p_2)^2 \approx -t = -(p'_1-p_1)^2 ~.
\end{equation}
\end{itemize}

Then,
replacing in (\ref{a}) the lepton-antilepton pair by $e^{-},~e^{+}$ and
the quark-antiquark pair by $\mu^{-},~ \mu^{+}$ respectively, we see
immediately  that the electric charge almost does not change its direction
 in the forward kinematics (\ref{smallt}) while it's reversed
 in the backward kinematics (\ref{smallu}) .  Obviously that
does not apply strictly for the annihilation into quarks because
the electric
charges of $u$ -quarks and $d$ -quarks are different in sign.
Therefore $t$ -kinematics is "forward" for the annihilation
into a $d\bar{d}$ -pair and at the same moment it is "backward"
 for the annihilation into $u\bar{u}$ quarks. We will come back
to this definition of backward and forward kinematics later,  when we shall
discuss the annihilation into $u~ \bar{u}$ and $d~
\bar{d}$ pairs, but  until then we refer to (\ref{smallt}) as
$t$ - kinematics, and (\ref{smallu}) as $u$ - kinematics.

In order to simplify the isospin structure, it is convenient to expand
the matrix $\widetilde{M}_{kk'}^{ii'}$ into a sum,
each term  corresponding to some
irreducible representation of $SU(2)$.
In the $t$ - kinematics (\ref{smallt}), the initial $t$ - channel state
is $l^k(p_1)\bar{q}_{i'}(p'_1)$. Obviously,

\begin{equation}
\label{tstates}
l^k(p_1)\bar{q}_{i'}(p'_1) = \left[ \frac{1}{2}
\delta^k_{i'}\delta^b_a  +
(\delta^k_a \delta^b_{i'} - \frac{1}{2}\delta^k_{i'} \delta^b_a )
\right] l^a(p_1)\bar{q}_{b}(p'_1) =
\left[ \frac{1}{2} \delta^k_{i'}\delta^b_a  +
2 \sum_c (t_c)^k_{i'}(t_c)^b_a \right] l^a(p_1)\bar{q}_{b}(p'_1)
\end{equation}

where the first term corresponds to the scalar and the second one -- to
the triplet representation of $SU(2)$.  Eq.~(\ref{tstates}) suggests the
the representation

\begin{equation}
\label{ainvt}
\widetilde{M}_{kk'}^{ii'} = (P_0)_{kk'}^{ii'} \widetilde{M}_0 +
(P_1)_{kk'}^{ii'} \widetilde{M}_1
\end{equation}

where $\widetilde{M}_{0,1}(s,~t)$ are scalar
functions and the singlet and triplet
projection operators correspondingly are :

\begin{equation}
\label{pt}
(P_0)_{kk'}^{ii'} = \frac12\delta_k^{i'} \delta^i_{k'},
 ~~~(P_1)_{kk'}^{ii'} = 2 (t_c)_k^{i}(t_c)^{i'}_{k'} ~.
\end{equation}

Similarly for the $u$ - kinematics (\ref{smallu}), where the
initial $u$ -channel state is $l^k(p_1) q^{k'}(-p'_2)$,  irreducible
$SU(2)$ -representations are obtained  by symmetrization and
antisymmetrization,

\begin{equation}
\label{ainvu}
\widetilde{M}_{kk'}^{ii'} = (P_-)_{kk'}^{ii'} \widetilde{M}_- (u, s) +
(P_+)_{kk'}^{ii'} \widetilde{M}_+ (u, s)~~,
\end{equation}

with

\begin{equation}
\label{pu}
(P_-)_{kk'}^{ii'} = \frac{1}{2} \left[\delta^i_k \delta^{i'}_{k'}-
\delta^{i'}_{k} \delta^i_{k'}\right]~,
~~~(P_+)_{kk'}^{ii'} = \frac{1}{2} \left[\delta^i_k \delta_{k'}^{i'} +
\delta^{i'}_{k} \delta^i_{k'}\right] ~.
\end{equation}

Using the projectors $P_j$ , $j = 0,1,-,+$, the invariant amplitudes
can be easily obtained:

\begin{equation}
\label{abornc}
\widetilde{M}_j = \frac{(P_j)^{kk'}_{ii'}(\widetilde{M})_{kk'}^{ii'}}
{(P_j)^{kk'}_{ii'}(P_j)_{kk'}^{ii'}} ~.
\end{equation}

In the Born approximation, the amplitudes $\widetilde{M}_j$ defined in
Eqs.~(\ref{ainvt},\ref{ainvu}) can be written as

\begin{eqnarray}
\label{aborn}
&&\widetilde{M}_j^{Born} = R \, A^{Born}_j(s)~,
~~~~A^{Born}_j(s)=\frac{s}{s+\imath\epsilon} \, a_j ~,\nonumber\\
&&R = \frac{\bar{u}(-p_2)[(1\!-\!\gamma_5)/2]
\gamma_{\nu}[(1\!+\!\gamma_5)/2]u(p_1)
\bar{u}(p'_1)[(1\!-\!\gamma_5)/2]\gamma_{\nu}
[(1\!+\!\gamma_5)/2]u(-p'_2)}{s}
\end{eqnarray}

where $R$ denotes the normalized spinor factor and $A^{Born}_j$
are scalar functions of $s$, differing only in constant group
factors $a_j$~.

As we discuss the particular case of the left fermions
we can drop the factors $[(1\pm\gamma_5)/2]$ and use the
following definition:

\begin{equation}
\label{r}
R = \frac{\bar{u}(-p_2)\gamma_{\nu}u(p_1)
\bar{u}(p'_1)\gamma_{\nu}u(-p'_2)}{s}  ~.
\end{equation}

For the left particles, the lepton and the quark
hypercharges are $Y_l=-1$ and $Y_q=1/3$ respectively.  The group
factors $a_j$ for the Born amplitudes are

\begin{equation}
\label{aj}
a_0=\frac{3g^2+{g'}^2 Y_lY_q}{4}, ~~a_1=\frac{-g^2+{g'}^2 Y_lY_q}{4},
~~a_-=\frac{- 3g^2 + {g'}^2 Y_lY_q}{4},~~a_+=\frac{g^2 + {g'}^2 Y_lY_q}{4}~~.
\end{equation}

The contributions proportional to $Y_lY_q$ in Eq.~(\ref{aj}) come
from the Born graph where the lepton line is connected to the quark one
by the $B$ -field, the other contributions come from the Born graphs with
propagators of $W_i$  -fields.  Accounting for all DL corrections
transforms
the coefficients $a_j$ into invariant amplitudes $M_j$,

\begin{equation} \label{minv}
\widetilde{M}_j = R \, M_j(s,u,t) ~,
\end{equation}

where, in DLA, $M_j$ depend on $s$, $t$, $u$ through logarithms. When
$M_j$ are
calculated,  Eqs.~(\ref{pt},\ref{pu}) allow us to immediately
express  the amplitudes $e^+e^-\to u\bar{u}$  and
$e^+e^-\to d\bar{d}$ in terms of $M_j$:

\begin{eqnarray}
\label{eet}
M(e^+e^-\to u\bar{u}) &=&  R \, M_1(s,t) ,  \nonumber\\
M(e^+e^-\to d\bar{d}) &=&  R \, \left[M_0(s,t)+M_1(s,t)\right]/2 ~,
\end{eqnarray}

for the annihilation in $t$ -kinematics (\ref{smallt}). Similarly,

\begin{eqnarray}
\label{eeu}
M(e^+e^-\to u\bar{u}) &=& R \, \left[M_-(s,t)+M_+(s,t)\right]/2 ~,
\nonumber\\
M(e^+e^-\to d\bar{d}) &=& R \, M_+(s,t) ~,
\end{eqnarray}

for the annihilation in $u$ -kinematics (\ref{smallu}).

\section{Evolution equations for the invariant amplitudes $M_j$.}

In this section we calculate $M_j$ in the high energy limit, by
constructing and
solving an IREE for them, as a generalization of the evolution equations
derived earlier in QCD. This approach
exploits the evolution of scattering amplitudes with respect to the
infrared cut-off $\mu$ in the transverse momentum space. Transverse
momenta of all virtual particles are supposed to obey

\begin{equation}
\label{mu}
k_{i\perp} > \mu, ~~~~~~~k_{i\perp} \perp p_1, p_2.
\end{equation}

The value of the cut-off $\mu$ must not be smaller than any
of the involved masses, otherwise it is arbitrary. Introducing  $\mu$
makes also possible to neglect the masses of all involved quarks and to
restrict ourselves to consider the evolution of
$M_j$ with respect to $\mu$ only. Then one can
take in the final formulas $\mu$ of order of the largest mass involved. In
DLA we can also neglect the difference between the masses of EW bosons
$M_W$ and $M_Z$, putting in the final expressions

\begin{equation}
\label{mwz}
\mu = M = M_Z \approx M_W ~.
\end{equation}

First we consider the annihilation in $t$ -kinematics  and construct
the IREE for $M_j$ with $j = 0,1$. According to
(\ref{smallt}), $t$ is small compared to $u , -s$. To bound it from
below we assume that

\begin{equation}
\label{bigt}
s \gg -t \gg \mu^2 ~.
\end{equation}

The main idea of the IREE consists in evoluting the
invariant amplitudes with respect to the infrared
cut-off $\mu$ by applying to them the differential operator

\begin{equation}
\label{dif}
-\mu^2 \partial /\partial \mu^2~,
\end{equation}

in the form

\begin{equation}
\label{lhs}
-\mu^2 \frac{\partial M_j}{\partial \mu^2}  =
\frac{\partial M_j}{\partial \rho} + \frac{\partial M_j}{\partial
\eta}~, \end{equation}

where we use  $u\approx -s$ in
this forward kinematics and have introduced the notations

\begin{equation}
\label{rhoeta}
\rho = \ln(s/ \mu^2), ~~~~~~\eta = \ln(-t/ \mu^2) ~.
\end{equation}

\begin{figure}[h]
\begin{center}
\begin{picture}(320,360)
\put(0,0){
\epsfbox{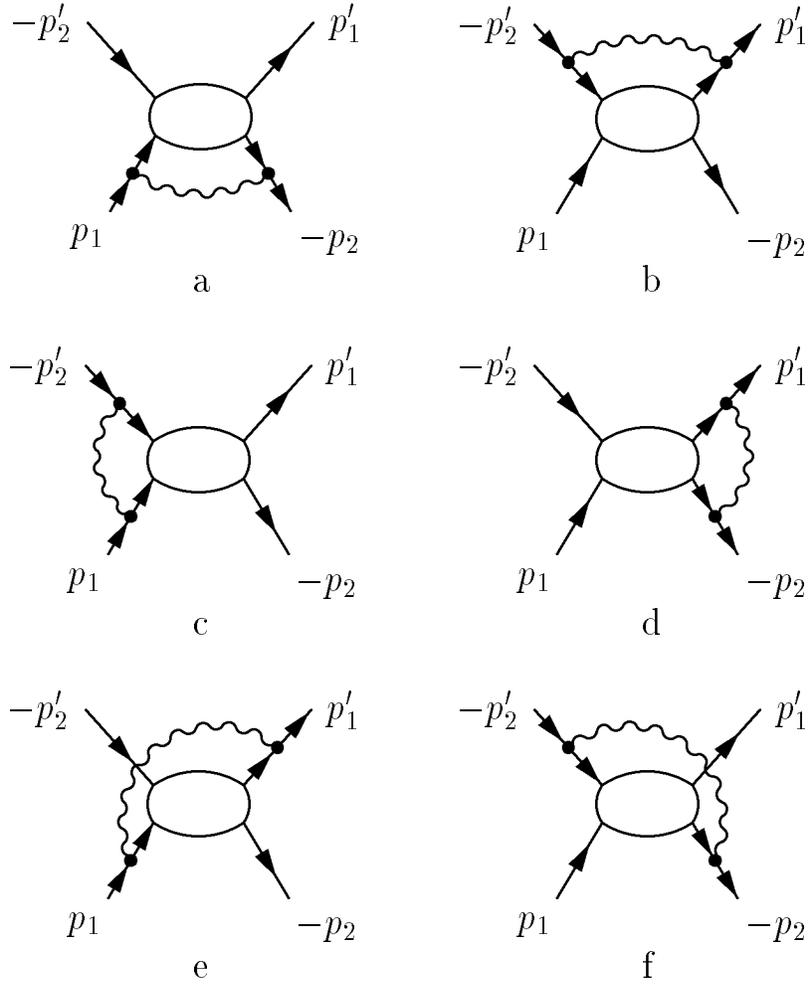}
}
\end{picture}
\end{center}
\caption{Contribution to IREE from soft EW boson factorization in
different channels: $s$-channel -- a and b, $u$-channel -- c and d,
$t$-channel -- e and f.}
\label{fig1}
\end{figure}

In order to obtain the rhs of last equation, we have to take into account
the
factorization
of DL contributions of virtual particles with respect to $\mu$, where
$\mu$ is the
lowest limit of integration over
$k_{\perp}$. In turn, this minimal $k_{\perp}$ acts as a
new cut-off for other virtual momenta (see \cite{kl,el} for
details).  When the virtual particle with the minimal $k_{\perp}$ is a
EW boson, one can factorize its DL contributions as shown in
Fig.~\ref{fig1}.  Applying  then the  differentiation (\ref{dif})
and the projection operators $P_j$ we
obtain with the help of Eq.~(\ref{abornc}) the contributions $G_0$, $G_1$
to the EW singlet and triplet parts of the IREE respectively.

Before  writing the explicit expressions for $G_{0,1}$, we want to discuss
the general structure of these contributions.  Integration over
longitudinal momentum of the factorized boson with momentum $k$ in
graphs (a) and (b) in Fig,~\ref{fig1} yields $\ln(-s/ k^2_{\perp})$
whereas the same integration in graphs (c) and (d) yields
$\ln(-u/k^2_{\perp}) \approx \ln(s/k^2_{\perp}) $.  Similarly, graphs
(e),(f) yield $\ln(-t/ k^2_{\perp})$.  Besides these logarithms, each
graph in Fig.~\ref{fig1} contains also an integration  of
$M_j(s/k^2_{\perp},t/k^2_{\perp}) /k^2_{\perp}$ over $k^2_{\perp}$,  with
the lowest limit $\mu^2$. After differentiation (\ref{dif}), we arrive
at

\begin{equation}
\label{ggeneral}
G_{0,1} = \frac{1}{8\pi^2}
\sum_{j=0,1} \left[ \left(b^{(j)}_s\right)_{0,1}
\ln\left(\frac{-s}{\mu^2}\right) -
\left(b^{(j)}_u\right)_{0,1} \ln\left(\frac{-u}{\mu^2}\right) \right]
M_j\left(\frac{s}{\mu^2},\frac{t}{\mu^2}\right) - \frac{1}{8\pi^2}
h_{0,1}
\ln\left(\frac{-t}{\mu^2}\right)
M_{0,1}\left(\frac{s}{\mu^2},\frac{t}{\mu^2}\right)~,
\end{equation}

%%%%
%%%%\begin{eqnarray}
%%%%\label{ggeneral}
%%%%G_0  &=& (1/8\pi^2)\left[b^{(0)}_s \ln(-s/ \mu^2) -
%%%%b^{(0)}_u \ln(-u/ \mu^2) - h_0 \ln(-t/\mu^2) \right]
%%%%M_0(s/\mu^2, t/\mu^2) +  \\ \nonumber
%%%%& &(1/4\pi^2)\left[b^{(1)}_s \ln(-s/ \mu^2)  -b^{(1)}_u \ln(-u/ \mu^2)
%%%%\right] M_1(s/\mu^2, t/\mu^2) ~.
%%%%\end{eqnarray}
%%%%

where the the quantities $\left(b^{(j)}_s\right)_{0,1}, ..., h_{0,1}$ will be
explicitly given later.

Let us introduce

\begin{equation}
\label{rhopm}
\rho^{(\pm)} = \frac{1}{2}\left [ \ln\left(\frac{-s}{\mu^2}\right)
\pm \ln\left(\frac{-u}{\mu^2}\right) \right]
\end{equation}

so that $\ln(-s/\mu^2)=\rho^{(+)}+\rho^{(-)}$ and
$\ln(-u/\mu^2)=\rho^{(+)}-\rho^{(-)}$.
Obviously,  $\rho^{(+)}$ and $\rho^{(-)}$ are symmetrical
and antisymmetrical functions with respect to replacing $s$ by $u$. It
is convenient also to introduce the invariant amplitudes
$M_{0,1}^{(\pm)}$ with the same properties
\footnote{In the Regge theory, amplitudes $M^{(\pm)}$ are called the
positive and negative signature amplitudes, and we use these notation
below.}
:

\begin{equation}
\label{mpm}
M^{(\pm)}_{0,1} = \frac12 \left[
M_{0,1}\left(\frac{s}{\mu^2},\frac{t}{\mu^2}\right) \pm
M_{0,1}\left(\frac{u}{\mu^2},\frac{t}{\mu^2}\right) \right]~,
\end{equation}

so that $M_{0,1} = M^{(+)}_{0,1}+M^{(-)}_{0,1}$.
Then for signature ampltudes $G_{0,1}^{(\pm)}$ defined as

\begin{equation}
\label{gpm}
G^{(\pm)}_{0,1} = \frac12\left[G_{0,1}(s,t) \pm G_{0,1}(u,t)\right]
\end{equation}

from Eq.~(\ref{ggeneral}) we obtain the following expressions

\begin{eqnarray}
\label{grho}
G^{(+)}_{0,1} &=& \frac{1}{8\pi^2} \sum_{j=0,1} \left[
\left(b^{(j)}\right)^{(+)}_{0,1} \rho^{(+)} M_j^{(+)} +
\left(b^{(j)}\right)^{(-)}_{0,1} \rho^{(-)} M_j^{(-)} \right]
+ \frac{1}{8\pi^2}\, h_{0,1}\, \eta\, M_{0,1}^{(+)}~, \nonumber\\
G^{(-)}_{0,1} &=& \frac{1}{8\pi^2} \sum_{j=0,1} \left[
\left(b^{(j)}\right)^{(+)}_{0,1} \rho^{(+)} M_j^{(-)} +
\left(b^{(j)}\right)^{(-)}_{0,1} \rho^{(-)} M_j^{(+)} \right]
+ \frac{1}{8\pi^2}\, h_{0,1}\, \eta\, M_{0,1}^{(-)}
\end{eqnarray}

%%%
%%%\begin{eqnarray}
%%%\label{grho}
%%%G_0^{(+)}  &=&
%%%(1/8\pi^2) \left[ b^{(+)}_0\rho^{(+)}M_0^{(+)} +
%%%b^{(-)}_0 \rho^{(-)} M_0^{(-)}  +  h_0 \eta M_0^{(+)}
%%%b^{(+)}_1 \rho^{(+)} M_1^{(+)} +
%%%b^{(-)}_1\rho^{(-)} M_1^{(-)} +  \right] \nonumber\\
%%%G_0^{(-)} &=& (1/8\pi^2)\left[(b^{(+)}_0 \rho^{(+)}M_0^{(-)}  +
%%%h_0 \eta M_0^{(-)} +
%%%b^{(-)}_0 \rho^{(-)} M_0^{(+)}  +
%%%b^{(+)}_1 \rho^{(+)}M_1^{(-)}  +
%%%b^{(-)}_1 \rho^{(-)}M_1^{(+)} \right] ~.
%%%\end{eqnarray}
%%%

where

\begin{equation}
\label{bpm}
\left(b^{(j)}\right)^{(\pm)}_{0,1} =
\left(b^{(j)}_s\right)_{0,1} \mp
\left(b^{(j)}_u\right)_{0,1}~.
\end{equation}

%%%
%%%\begin{equation}
%%%\label{bpm}
%%%b^{(\pm)}_0 \equiv b^{(0)}_s \mp b^{(0)}_u,~~~~~
%%%b^{(\pm)}_1 \equiv b^{(1)}_s \mp b^{(1)}_u
%%%\end{equation}
%%%

\begin{figure}
\begin{center}
\begin{picture}(370,130)
\put(0,0){
\epsfbox{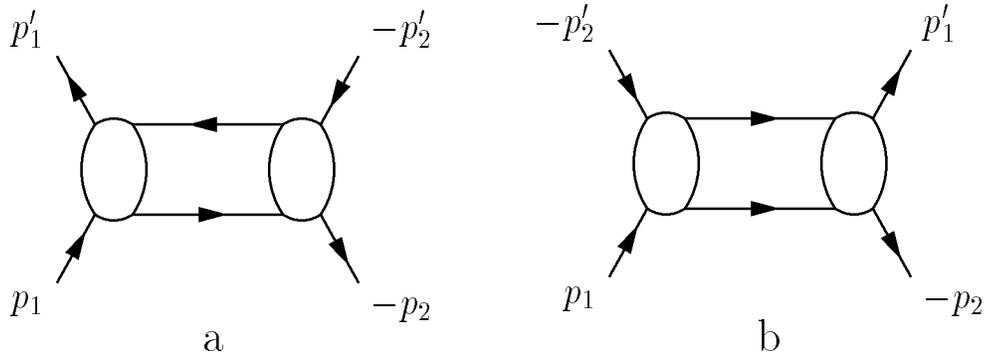}
}
\end{picture}
\end{center}
\caption{Contribution to IREE from soft fermion intermediate state in
$t$-channel -- a and in $u$-channel -- b.}
\label{fig2}
\end{figure}

Besides an EW boson, in kinematics (\ref{smallt}) a $t$ -channel virtual
fermion pair, as shown in Fig.~\ref{fig2}a,  could also attain the minimal
transverse momentum.  However, DL contributions arising from
the integration over
this pair momentum could only come from the region
$k^2_{\perp} > -t \gg \mu^2$.  Hence they do not depend on $\mu$ in
kinematics (\ref{bigt}) and must vanish when differentiated with
respect to $\mu$.  The same is true for the Born amplitudes
(\ref{aborn}).

As soon as $\ln(-s/\mu^2)=\ln(s/\mu^2)-\imath\pi$,
in kinematics (\ref{smallt}) with
$\rho=\ln(s/\mu^2)\approx\ln(-u/\mu^2)$ we obtain

\begin{eqnarray}
\label{logpm}
\rho^{(+)} &=&\rho - \frac{\imath\pi}{2}\,\mbox{sign}(s) ~,
\nonumber\\
\rho^{(-)} &=& -\frac{\imath\pi}{2}\,\mbox{sign}(s) ~.
\end{eqnarray}

It is assumed in DLA  that $\ln(s/\mu^2)\gg\pi$. This means that

\begin{equation}
\label{rhoprhom}
\rho^{(+)} \gg \rho^{(-)}   ~.
\end{equation}

Similarly, in DLA in each order of the perturbative expansion the
amplitudes
$M^{(+)}_j$ dominate over $M^{(-)}_j$ for one power of
$\ln(s/\mu^2)$.
By the same reason the amplitudes $M^{(+)}_j$ are mainly real,
and we can assume

\begin{equation}
\label{mpmm}
M^{(+)}_j\approx \Re M^{(+)}_j \gg \left|M^{(-)}_j\right| ~.
\end{equation}

Combining Eqs.~(\ref{lhs},\ref{grho}) and using
Eqs.~(\ref{mpmm},\ref{logpm},\ref{rhoprhom}) leads us to the following
IREE where negligible in DLA terms $M^{(-)}\rho^{(-)}$
are dropped and terms $M^{(-)}\rho^{(+)}\sim M^{(+)}\rho^{(-)}$ are
retained:

%%%
%%%\begin{eqnarray}
%%%\label{gpmapprox}
%%%G_{0,1}^{(+)}  &=& \frac{1}{8\pi^2}\, \rho\, \left[ \sum_{j=0,1}
%%%\left(b^{(j)}\right)^{(+)}_{0,1} M_j^{(+)} \right] +
%%%\frac{1}{8\pi^2}\, h_{0,1}\, \eta\, M_{0,1}^{(+)}\\ \nonumber
%%%G_{0,1}^{(-)} &=& \frac{1}{8\pi^2}\, \rho\, \left[ \sum_{j=0,1}
%%%\left(b^{(j)}\right)^{(+)}_{0,1} M_j^{(-)} \right] +
%%%\left( \frac{-\imath\pi}{2}\right)\,
%%%\frac{1}{8\pi^2}\, \left[ \sum_{j=0,1}
%%%\left(b^{(j)}\right)^{(-)}_{0,1} M_j^{(+)} \right] +
%%%\frac{1}{8\pi^2}\, h_{0,1}\, \eta\, M_{0,1}^{(-)}~.
%%%\end{eqnarray}

%%%
%%%\begin{eqnarray}
%%%\label{gpmapprox}
%%%G_0^{(+)}  &=&
%%%(1/8\pi^2) \rho \left[ b^{(+)}_0 M_0^{(+)} +
%%%b^{(+)}_1 M_1^{(+)} \right] + (1/8\pi^2) h_0 \eta M_0^{(+)}\\ \nonumber
%%%G_0^{(-)} &=&
%%%(1/8\pi^2) \rho \left[ b^{(+)}_0 M_0^{(-)}  +
%%%b^{(+)}_1 M_1^{(-)} \right] +  (-\imath \pi/2)(1/8\pi^2)
%%%\left[ b^{(-)}_0 M_0^{(+)} + b^{(-)}_1 M_1^{(+)} \right]
%%%+ (1/8\pi^2) h_0 \eta M_0^{(+)}~.
%%%\end{eqnarray}
%%%

\begin{eqnarray}
\label{m01}
\frac{\partial M_{0,1}^{(+)}}{\partial\rho} +
\frac{\partial M_{0,1}^{(+)}}{\partial\eta} &=&
\frac{1}{8\pi^2}\, \rho\, \left[ \sum_{j=0,1}
\left(b^{(j)}\right)^{(+)}_{0,1} M_j^{(+)} \right] +
\frac{1}{8\pi^2}\, h_{0,1}\, \eta\, M_{0,1}^{(+)}~,\\ \nonumber
\frac{\partial M_{0,1}^{(-)}}{\partial \rho} +
\frac{\partial M_{0,1}^{(-)}}{\partial \eta} &=&
\frac{1}{8\pi^2}\, \rho\, \left[ \sum_{j=0,1}
\left(b^{(j)}\right)^{(+)}_{0,1} M_j^{(-)} \right] +
\left( \frac{-\imath\pi}{2}\right)\,
\frac{1}{8\pi^2}\, \left[ \sum_{j=0,1}
\left(b^{(j)}\right)^{(-)}_{0,1} M_j^{(+)} \right] +
\frac{1}{8\pi^2}\, h_{0,1}\, \eta\, M_{0,1}^{(-)}~.
\end{eqnarray}

%%%
%%%\begin{eqnarray}
%%%\label{m01}
%%%\frac{\partial M_0^{(+)}}{\partial\rho} +
%%%\frac{\partial M_0^{(+)}}{\partial\eta} &=&
%%%(1/8\pi^2) \rho \left[ b^{(+)}_0 M_0^{(+)} +
%%%b^{(+)}_1 M_1^{(+)} \right] + (1/8\pi^2) h_0 \eta M_0^{(+)}\\ \nonumber
%%%\frac{\partial M_0^{(-)}}{\partial \rho} +
%%%\frac{\partial M_0^{(-)}}{\partial \eta} &=&
%%%(1/8\pi^2) \rho \left[ b^{(+)}_0 M_0^{(-)}  +
%%%b^{(+)}_1 M_1^{(-)} \right] +  (-\imath \pi/2)(1/8\pi^2)
%%%\left[ b^{(-)}_0 M_0^{(+)} + b^{(-)}_1 M_1^{(+)} \right]
%%%+ (1/8\pi^2) h_0 \eta M_0^{(+)}~.
%%%\end{eqnarray}
%%%

Let us proceed now to the  $u$ - kinematics (\ref{smallu}).
Using projection operators (\ref{pu}) instead of (\ref{pt})
one can consider the annihilation in the $u$ -kinematics and obtain
the IREE for invariant signature amplitudes
$M^{(\pm)}_-$, $M^{(\pm)}_+$ introduced in a way  similar to
that one used for $M^{(\pm)}_{0,1}$.  As the amplitudes  $M^{(\pm)}_-$,
$M^{(\pm)}_+$ correspond to SU(2) singlet and triplet representations,
similarly to $M^{(\pm)}_{0,1}$, we can easily obtain IREE for  $u$
-kinematics from Eq.~(\ref{m01}) with the replacement

\begin{equation}
\label{ttoureplace}
t\longleftrightarrow u~,\quad "0"\to "-"~,\quad "1"\to "+" ~,\quad
Y_q\to -\,Y_q ~.
\end{equation}

Indeed, adding the restriction

\begin{equation}
\label{bigu}
s \gg -u \gg \mu^2 ~
\end{equation}

to Eq.~(\ref{smallu}), the derivation is quite similar to the
previous one done for the $t$
- kinematics thus leading to the same structure of the IREE for the
amplitudes $M^{(\pm)}_{-,+}$.
Therefore one can write down the same IREE for all invariant signature
amplitudes  $M_j^{(\pm)}$, with $j = 0,1,"-","+"$,
generalizing Eq.~(\ref{m01}).

The next significant simplification of (\ref{m01}) comes
from the explicit calculation of the group factors
$\left(b^{(j)}\right)^{(\pm)}_{0,1}$. It turns out that

\begin{equation}\label{nondiagb}
\left(b^{(0)}\right)^{(+)}_1 = \left(b^{(1)}\right)^{(+)}_0 =
\left(b^{(-)}\right)^{(+)}_+ = \left(b^{(+)}\right)^{(+)}_- = 0,
\end{equation}

and consequently the  IREE for the positive signature  amplitudes become
linear
homogenous partial differential equations.  They can be written in a
more simple general way:

\begin{equation}
\label{positive}
\frac{\partial M_j^{(+)}}{\partial \rho} +
\frac{\partial M_j^{(+)}}{\partial \eta'} =
-\, \frac{1}{8\pi^2} \left[ b_j\rho + h_j\eta' \right] M_j^{(+)}~,
\end{equation}

where  $b_j,  b_j$, and $\eta'$  will be specified below, so that the
only
difference between the equations for different amplitudes
comes from the numerical factors $b_j$, $h_j$:

\begin{eqnarray}
\label{bj}
b_0 = \frac{{g'}^2(Y_l - Y_q)^2}{4} ,\qquad
b_1 = \frac{8 g^2 + {g'}^2 (Y_l-Y_q)^2}{4} ,  \\  \nonumber
b_- = \frac{{g'}^2(Y_l+Y_q)^2}{4} ,\qquad
b_+ = \frac{8 g^2 + {g'}^2 (Y_l+Y_q)^2}{4} ,
\end{eqnarray}

\begin{eqnarray}
\label{hj}
h_0 = \frac{3 g^2 + {g'}^2 Y_l Y_q}{2} ,\qquad
h_1 = \frac{-g^2 + {g'}^2 Y_l Y_q}{2} ,  \\  \nonumber
h_- = \frac{3 g^2 - {g'}^2 Y_l Y_q}{2} ,\qquad
h_+ = \frac{-g^2 - {g'}^2 Y_l Y_q}{2} .
\end{eqnarray}

The IREE for the negative signature amplitudes $M_j^{(-)}$
are also linear partial differential equations,  but not being homogenous,
they involve
positive signature amplitudes through a non-zero matrix $r_{jj'}$:

\begin{equation}
\label{negative}
\frac{\partial M_j^{(-)}}{\partial \rho} +
\frac{\partial M_j^{(-)}}{\partial \eta'} =
-\, \frac{1}{8\pi^2}\, \left[
\left( b_j\rho+h_j\eta'\right)\, M_j^{(-)} +
\left(\frac{-\imath\pi}{2}\right)\,\sum_{j'} r_{jj'}\,
M_{j'}^{(+)}\right]~.
\end{equation}

In order to write down the IREE for all $M_j^{(-)}$
and $M_j^{(+)}$ in the same way, we
have used in Eqs.~(\ref{positive},\ref{negative})
the variable $\eta'$  so that

 \begin{equation}
\label{etat}
\eta' \equiv \eta = \ln(-t/\mu^2)
\end{equation}

for $t$ - kinematics and

\begin{equation}
\label{etau}
\eta' \equiv \chi = \ln(-u/\mu^2)
\end{equation}

for $u$ - kinematics.

The non-zero numerical factors $r_{jj'}$ in Eq.~(\ref{negative})
are :

\begin{equation}
\label{rcc}
r_{00} = r_{11} = {g'}^2\, \frac{(Y_l + Y_q)^2}{4}~,\quad
r_{01} = r_{-+} = 3 g^2~, \quad
r_{10} = r_{+-} = g^2~,\quad
r_{--} = r_{++} = {g'}^2\, \frac{(Y_l - Y_q)^2}{4} ~.
\end{equation}

%%%All nonzero factors $a_j, ~b_j,  ~h_j, ~r_{jj'}$  are listed in
%%%Table~1.
%%%
%%%
%%%
%%%          Table 1.
%%%
%%%
%%%The common feature for kinematics (\ref{bigt}) and (\ref{bigu}) is
%%%that in each of the kinematics the produced quarks, in cmf,
%%%are concentrated
%%%in the narrow cone with opening angle
%%%
%%%\begin{equation}
%%%\label{thetatu}
%%%1\gg\theta\gg\theta_0\equiv\sqrt{M^2/s}
%%%\end{equation}
%%%
%%%around the initial lepton-antilepton beam.
%%%Until we consider flavours of the
%%%produced quarks, we can treat scattering amplitudes $M_j^{(\pm)}$
%%%with $j = 0,1,-,+$ in kinematics (\ref{bigt}),(\ref{bigu}) the same way.
%%%

\section{Solutions to IREE for the invariant amplitudes $M_j^{(\pm)}$.}

It is easy to check that a general solution to
Eq.~(\ref{positive}) is

\begin{equation}
\label{mposgen}
M^{(+)}_j = \Phi^{(+)}_j(\rho - \eta')\,\exp[-\phi_j(C,\,\eta')]~,
\end{equation}

with

\begin{equation}
\label{phi}
\phi_j = \frac{1}{8\pi^2}\,
\left[ b_j\,C\,\eta'+ (b_j + h_j)\, \frac{{\eta'}^2}{2} \right]~,
\end{equation}

$C=\rho-\eta'=\mbox{const}$ and $\Phi_j$ is an arbitrary
function.  We can specify it by imposing the boundary condition

\begin{equation}
\label{borderpos}
\left. M^{(+)}_j(\rho,\eta')\right|_{\eta' = 0} = A_j^{(+)}(\rho) ~,
\end{equation}

where $A_j^{(+)}(\rho)$ are the amplitudes for the annihilation in
the ``collinear kinematics'', i.e. in the kinematics where quarks are
produced close to the direction of the beam of initial
leptons, with the value of either $t$ or $u$ much smaller than
those fixed by Eqs.~(\ref{bigt},\ref{bigu}).
Of course one must use a separate boundary condition
(\ref{borderpos}) for the
$t$ and $u$ - kinematics. We define the notation ``collinear $t$
-kinematics'' for

\begin{equation}
\label{colt}
-t < \mu^2
\end{equation}

and the notation ``collinear $u$ -kinematics'' for

\begin{equation}
\label{colu}
-u < \mu^2 ~.
\end{equation}

It is convenient to use the Mellin transform to represent
signature amplitudes $A^{(\pm)}_j$ in the ``collinear kinematics''
(\ref{colt},\ref{colu}):

\begin{equation}
\label{mellin}
A^{(\pm)}_j(\rho) =
\int_{-\imath \infty}^{\imath \infty}
\frac{d\omega}{2\pi\imath} \left(\frac{s}{\mu^2}\right)^{\omega}\,
\xi^{(\pm)}(\omega)\,
F_j^{(\pm)}(\omega)
\end{equation}

where

\begin{equation}
\label{sign}
\xi^{(\pm)} = \frac{\exp(-\imath\pi\omega) \pm 1}{2}
\end{equation}

are the well known signature factors. At asymptotically high energy $s$
the region of small $\omega$, $\omega\ll 1$, is dominating in
integral (\ref{mellin}). This allows one to exploit the following
approximations:

\begin{equation}
\label{signas}
\xi^{(+)}\approx 1~,\qquad \xi^{(-)}\approx -\,\frac{\imath\pi\omega}{2}
\end{equation}

Eq.~(\ref{mellin}) implies that
the positive signature amplitudes  $M^{(+)}_j(\rho, \eta')$ in
the kinematic regions (\ref{bigt},\ref{bigu}) can be easily
expressed through the Mellin amplitudes $F_j^{(\pm)}(\omega)$ :

\begin{equation}
\label{mpos}
M^{(+)}_j(\rho, \eta') = \exp[-\phi_j(C,\, \eta')]
\int_{-\imath\infty}^{\imath\infty}
\frac{d\omega}{2\pi\imath}\, e^{(\rho-\eta')\omega}\,
\xi^{(+)}(\omega)\, F_j^{(+)}(\omega) ~.
\end{equation}

On the other hand, as stated earlier,
the IREE (\ref{negative}) for the
negative signature amplitudes  $M^{(-)}_j$  are not homogeneous, in
contrast to Eq.~(\ref{positive})
Besides the amplitudes $M^{(-)}_j$, they also involve the positive
signature amplitudes $M^{(+)}_j$.  In order to solve
Eq.~(\ref{negative}), we have to use again the boundary condition

\begin{equation}
\label{borderneg}
\left. M^{(-)}_j(\rho,\eta')\right|_{\eta' = 0} = A_j^{(-)}(\rho) ~.
\end{equation}

It is easy to check that the solution to the Eq.~(\ref{negative})
satisfying
Eq.~(\ref{borderneg})  is

\begin{equation}
\label{mneg}
M^{(-)}_j(\rho, \eta') = \exp[-\phi_j(C,\eta')]\, \left[
A_j^{(-)}(\rho-\eta') -
\int_{0}^{\eta'} d\tau\, \exp[\phi_j(C, \tau)]\,
\left(\frac{-\imath\pi}{2}\right)\, \left(\frac{1}{8\pi^2}\right)\,
\sum_{j'} r_{jj'}\, M^{(+)}_{j'}(C,\tau) \right] ~.
\end{equation}

Applying the Mellin transforms (\ref{mellin}) and (\ref{mpos}) for the
amplitudes in Eq.~(\ref{mneg})  we can rewrite it through the Mellin
amplitudes $F_j^{(\pm)}(\omega)$ as well.  In order to simplify this
procedure we use the substitution

\begin{equation}
\label{posnegsign}
\left(\frac{-\imath\pi}{2}\right)\, \xi^{(+)} \approx
\frac{1}{\omega}\,\xi^{(-)},
\end{equation}

which follows from the approximation (\ref{signas}) and is reasonable
when the small $\omega$ region is dominating in the integral.
Eventually we arrive at

\begin{eqnarray}
\label{mnegf}
M^{(-)}_j(\rho, \eta') &=& \exp[-\phi_j(C,\eta')]\,
\int_{-\imath\infty}^{\imath\infty}
\frac{d\omega}{2\pi\imath}\, e^{\omega(\rho-\eta')}\,
\xi^{(-)}(\omega)\, \left[ F_j^{(-)}(\omega)\,\, - \right.
\nonumber\\ & & \frac{1}{8\pi^2}\,
\sum_{j'} r_{jj'}\, \frac{F_{j'}^{(+)}(\omega)}{\omega}
\left. \int_0^{\eta'} d\tau\,
\exp[\phi_j(C,\tau)-\phi_{j'}(C,\tau)]\, \right] ~.
\end{eqnarray}

Eqs.~(\ref{mpos},\ref{mnegf})  show how one can obtain the
amplitudes  $M^{(\pm)}_j$ when the Mellin
amplitudes $F_j^{(\pm)}(\omega)$  are calculated.

\section{IREE for the Mellin amplitudes $F^{(\pm)}_j$ ~.}

In order to calculate $F^{(\pm)}_j$  we have to
construct IREE for ``collinear kinematics ''amplitudes
$A^{(\pm)}_j(\rho)$.  The IREE  for them differ from the
IREE for amplitudes $M^{(\pm)}_j$ considered in the previous section,
by the following reasons.

\begin{itemize}

\item[(i)] The amplitudes  $A^{(\pm)}_j$ in the kinematics
$(\ref{colt},\ref{colu})$ depend on $\rho = \ln(s/\mu^2)$ only,
and some graphs in Fig.~1 do not yield DL contributions to
IREE for them. In particular, graphs (e) and (f) with
factorized $t$ - channel virtual bosons do not contribute to IREE for
$A^{(\pm)}_{0,1}$ while graphs (c),(d) with factorized $u$ - channel
bosons do not contribute to IREE for $A^{(\pm)}_{-,+}$. The lhs of the
IREE for $A^{(\pm)}_j$ turns to  $-\mu^2 \partial A^{(\pm)}_j /
\partial \mu^2$ that in terms of the Mellin variable $\omega$
results in $\omega F^{(\pm)}_j(\omega)$.

\item[(ii)] The DL contributions of the graphs in Fig.~\ref{fig2} also
depend on $\mu^2$  and therefore do not vanish when
differentiated with respect to $\mu$.
As these graphs are convolutions of two amplitudes, their contributions
become simpler after applying the Mellin transform
(\ref{mellin}).  Then the differentiation
$-\mu^2\partial/\partial\mu^2$ of these graphs leads to the following
contribution to the  IREE for  the Mellin transforms $F^{(\pm)}_j$~:

\begin{equation}
\label{softq}
\frac{c_j}{8\pi^2} \left[ F^{(\pm)}_j(\omega) \right]^2 ~,
\end{equation}

with $c_j = 1$ for $j=0,1$ and $c_j=-1$ for $j="-","+"$.

\item[(iii)] Though at the first sight the Born amplitudes  $A^{Born}_j$
of Eq.~(\ref{aborn}) do not depend on $\mu^2$ it is necessary to
replace them by $a_j s/(s-\mu^2+\imath\epsilon)$. This form explicitly
tells that $s$ cannot be smaller than $\mu^2$ and also it
makes the Mellin transform for the Born amplitudes to be correctly
defined.  The Mellin transforms for Born amplitudes are therefore
$a_j/\omega$.  Applying $-\mu^2\partial/\partial\mu^2$ to them results
in multiplying by $\omega$. Hence the contributions of the Born
amplitudes to IREE are just the constant terms $a_j$.

\end{itemize}

As a result we arrive to the following IREE for Mellin
amplitudes $F^{(\pm)}_j(\omega)$.

\begin{eqnarray}
\label{eqfpos}
\omega F^{(+)}_j(\omega) &=& a_j +
\frac{b_j}{8\pi^2} \frac{dF^{(+)}_j(\omega)}{d \omega} +
\frac{c_j}{8\pi^2} \left[F^{(+)}_j(\omega)\right]^2 ~, \\
\label{eqfneg}
\omega F^{(-)}_j(\omega) &=& a_j +
\frac{b_j}{8\pi^2}\frac{1}{\omega}
\frac{d \left(\omega F^{(-)}_j(\omega)\right)}{d \omega} +
\frac{c_j}{8\pi^2} \left[F^{(-)}_j(\omega)\right]^2  -
\sum_{j'} \frac{r_{jj'}}{8\pi^2} F^{(+)}_{j'}(\omega) ~.
\end{eqnarray}

The coefficients $a_j$, $b_j$, $c_j$ and $r_{jj'}$ are listed in
Table~\ref{table1} for $t$ - kinematics and in  Table~\ref{table2} for
$u$ - kinematics.

Solutions to Eq.~(\ref{eqfpos}) can be expressed
in terms of the Parabolic cylinder functions $D_p$:

\begin{equation}
\label{fpos}
F^{(+)}_j(\omega) = \frac{a_j}{\lambda_j}
\frac{D_{p_j - 1} (\omega/\lambda_j)}{D_{p_j} (\omega/\lambda_j)}
\end{equation}

where

\begin{equation}
\label{fp}
\lambda_j=\sqrt{\frac{b_j}{8\pi^2}}~,\qquad p_j=\frac{a_j c_j}{b_j}~.
\end{equation}

In contrast, solutions to Eq.~(\ref{eqfneg}) can be
found only numerically. In  QED the negative signature
amplitudes for the backward $e^+e^-\to\mu^+\mu^-$ annihilation were
solved in Ref.~\cite{gln}. It is interesting to note that $b_0 =
0$ in the IREE for the forward amplitudes $M_{e\mu}^{(\pm)}$ of $e^+e^-
\to
\mu^+ \mu^-$ -annihilation, and the differential equations
(\ref{eqfpos})  for  $M_{e\mu}^{(\pm)}$ in kinematics
(\ref{colt}) turn into purely algebraic equations. This result was first
obtained  in Ref.~\cite{gg}. Later it was proved\cite{kl} that
the IREE for the (colourless) scalar components of the
$SU(3)$ negative signature amplitudes of
quark-antiquark annihilation into another quark-antiquark pair are
also algebraic and therefore can be easily solved.
The processes mentioned above are the only known examples of solving
IREE for negative signature amplitudes
\footnote{In the context of the
EW theory, the IREE for the backward scattering amplitude with the
negative signature was solved in \cite{flmm} for the irrealistic case
of a complex value of the Weinberg angle.}.
In all those cases the
intercepts of the negative signature amplitudes are greater than those
for the positive signature amplitudes, though the difference amounts
only to a few percents.  Equations for the negative signature amplitudes
always involve the positive signature amplitudes. It has been
observed\cite{egt} in a QCD context, that these amplitudes can be
approximated by their Born values with good accuracy. Such an
approximation can help in
solving Eqs.~(\ref{eqfneg}). We do not consider explicit solutions of
Eqs.~(\ref{eqfneg}) in the present paper.  Instead, we consider below
only contributions of amplitudes with the positive signature
$M_j^{(+)}$.  Combining Eqs.~(\ref{mpos},\ref{fpos}) and introducing
variable $x = \omega/ \lambda_j$, we arrive at the expression

\begin{equation}
\label{mposit}
M^{(+)}_j(\rho, \eta') = a_j\, \exp[-\phi_j(\rho-\eta',\eta')]
\int_{-\imath\infty}^{\imath\infty} \frac{dx}{2\pi\imath}\,
e^{\lambda_j x(\rho-\eta')}\, \frac{D_{p_j-1}(x)}{D_{p_j}(x)} ~.
\end{equation}

It is useful here to split $\phi_j$ defined by Eq.~(\ref{phi}) and to
combine its part depending on $\rho-\eta'$ with the exponent of the
integrand in Eq.~(\ref{mposit}). Then changing the integration variable
$x$ to $l$, where $x=l+\lambda_j\eta'$ we finally obtain

\begin{equation}
\label{mpostu}
M^{(+)}_j(\rho,\eta') = a_j\,
\exp\left[-\,\frac{b_j+h_j}{8\pi^2}\,\frac{{\eta'}^2}{2}\right]
\int_{-\imath \infty}^{\imath \infty} \frac{dl}{2\pi\imath}\,
e^{\lambda_j l (\rho-\eta')}
\frac{D_{p_j-1}(l+\lambda_j\eta')}
{D_{p_j}(l+\lambda_j\eta')}
\end{equation}

where, for the case of $t$ - kinematics with $j=0,1$,
$\eta'=\eta=\ln(-t/\mu^2)$ and for the case of $u$ - kinematics with
$\eta'=\chi=\ln(-u/\mu^2)$, $j="-","+"$.

The exponential factor in front of the integral in Eq.~(\ref{mpostu}) is
of Sudakov type. Actually it is a product of the Sudakov form
factors of the left lepton and of the left
quark.  As follows from Eqs.~(\ref{bj},\ref{hj}), it does not depend on
$j$, i.e.  is same for all invariant amplitudes,

\begin{equation}
\label{sudakew}
S= \exp\left[-\,\frac{1}{8\pi^2} \,
\left(\frac32\, g^2 + \frac{Y_l^2+Y_q^2}{4}\, {g'}^2 \right)
\, \frac{{\eta'}^2}{2}\right] ~.
\end{equation}

It corresponds to DL contributions of soft virtual EW bosons and
vanishes in the final expressions for the cross sections when
bremsstrahlung of
soft EW bosons are taken into account. Assuming this to be done we can
omit
such Sudakov factors.

Until now we have discussed the annihilation $l\bar{l}\to q\bar{q}$
for the case when the both initial and final particles were left, i.e.
the spinors in Eq.~(\ref{r}) were actually $[(1+\gamma_5)/2] u$.
It is clear that applying the same reasoning it is easy to construct
IREE for amplitudes in $t$ and $u$ -kinematics with right fermions.
Solutions to such IREE can be presented in the same form
of Eq.~(\ref{mpostu})
with $j=RR$ for both right leptons and quarks, and $j=LR$ when
the initial leptons are left whereas the final quarks are right, or
$j=RL$ vice versa, with:

\begin{equation}
\label{right}
\begin{array}{llll}
a_{RR} = \frac{{g'}^2 Y_l Y_q}{4}~, &
\lambda_{RR} = \sqrt{\frac{b_{RR}}{8\pi^2}}~, &b_{RR} =
\frac{{g'}^2(Y_l \mp Y_q)^2}{4}~, & h_{RR} =
\pm\frac{{g'}^2\, Y_l Y_q}{2}~,\\
a_{LR} = \frac{{g'}^2 Y_l Y_q}{4}~, &
\lambda_{LR} = \sqrt{\frac{b_{LR}}{8\pi^2}}~, & b_{LR} =
\frac{3g^2 + {g'}^2(Y_l \mp Y_q)^2}{4}~, & h_{LR} =
\pm\frac{{g'}^2\, Y_l Y_q}{2} ~,
\end{array}
\end{equation}

where $\mp$ signs in $b$ and $\pm$ signs in $h$ correspond to $t$ and
$u$ - kinematics respectively.  It is worthwhile to remind here that in
the numerical estimations when using Eq.~(\ref{right})) one has to
substitute for $Y_l$, $Y_q$ for right and left fermions the correct EW
hypercharges:  $Y=2Q$ for right fermions and $Y=2(Q-T_3)$ for left
fermions.  The same formulae for the invariant
amplitude $M_{LR}^{(+)}$ in the collinear $u$ -kinematics, with $\chi=0$,
can be also obtained from results of Ref.~\cite{flmm}.

\section{Forward -- backward asymmetry}

In this section we consider in detail the
$e^+e^-$ - annihilation
into quark-antiquark pairs of  different flavours in the configuration of
forward and backward kinematics.
According to the terminology introduced in Ref.~\cite{gg}, also discussed
earlier in section~\ref{sect1}, the forward kinematics for
$e^+(p_1)e^-(p_2) \to u(p'_1)\bar{u}(p'_2)$ - annihilation
corresponds actually to the
$u$ - kinematics defined by  Eq.~(\ref{smallt}) because the electric
charges
of the electron and the $u$ - quark are opposite. On the contrary,
the forward
kinematics for the annihilation of $e^+e^- \to d \bar{d}$ corresponds
to the $t$ - kinematics, as
we have defined in  Eq.~(\ref{smallu}). Similarly,
 the backward kinematics for $e^+(p_1)e^-(p_2) \to u
(p'_1)\bar{u}(p'_2)$ -annihilation is a $t$ - kinematics and that
for $e^+e^- \to d \bar{d}$ is the $u$ - channel one. From the results
we have obtained in section~\ref{sect1} it follows that the amplitude for
forward $e^+e^- \to u\bar{u}$ -annihilation, $M^F_u$ is expressed in
terms of amplitudes $M_-, M_+$ of Eq.~(\ref{mpostu}):

\begin{equation}
\label{mfu}
M_u^F =  \frac{ M_- + M_+}{2} \approx  \frac{ M^{(+)}_- +
M^{(+)}_+}{2} ~,
\end{equation}

whereas the backward amplitude $M_u^B$
for the same quarks is expressed through the amplitude $M_1$ of
Eq.~(\ref{mpostu}):

\begin{equation}
\label{mbu}
M_u^B = M_1 \approx  M_1^{(+)}  ~.
\end{equation}

Similarly, the forward amplitude $M^F_d$ for
$e^+e^- \to d\bar{d}$ is

\begin{equation}
\label{mfd}
M_d^F =  \frac{M_0 + M_1}{2} \approx  \frac{ M^{(+)}_0 +
M^{(+)}_1}{2}  ~,
\end{equation}

while  the backward amplitude for this process is

\begin{equation}
\label{mbd}
M_d^B = M_+ \approx  M^{(+)}_+  ~.
\end{equation}

By forward kinematics for $e^+e^-\to q\bar{q}$ -annihilation
we mean that quarks with positive electric
charges, $u$ and $\bar{d}$, are produced around the initial $e^+$ - beam,
in the cmf,
within a cone with a small opening angle $\theta$,

\begin{equation}
\label{thetatu}
1\gg\theta\ge \theta_0 = \frac{2M}{\sqrt{s}}~
\end{equation}

By backward kinematics we means just the opposite -- the electric charge
scatters backwards in a cone with the same opening angles.

The differential cross section $d\sigma_F$ for the forward annihilation is

\begin{equation}
\label{sigmaf}
d\sigma_F = d\sigma^{(0)}\,\left[|M_u^F|^2 + |M_d^F|^2\right] \equiv
d\sigma^{(0)}\,\left[F_u + F_d\right]  ~,
\end{equation}

and similarly, the differential cross section $d\sigma_B$ for the
backward annihilation is

\begin{equation}
\label{sigmab}
d\sigma_B = d\sigma^{(0)}\, \left[|M_u^B|^2 + |M_d^B|^2\right] \equiv
d\sigma^{(0)}\, \left[B_u + B_d\right],
\end{equation}

where $d\sigma^{(0)}$ stands
for the Born cross section, though without couplings. It absorbs
the factor $|R|^2$ defined by Eq.~(\ref{r}) which cancels in the
expression for the forward-backward asymmetry which we define as

\begin{equation}
\label{defasym}
A \equiv \frac{d\sigma_F - d\sigma_B }{d\sigma_F + d\sigma_B}
= \frac{F - B}{F + B} ~
\end{equation}

where

\begin{equation}
\label{FB}
F = F_u + F_d, \qquad  B = B_u + B_d ~.
\end{equation}

%According to Eqs.~(\ref{mpostu},\ref{thetatu}) the cross sections $F$,
%$B$, and therefore the value of asymmetry $A$, depend on the angle
%$\theta$ when $\theta>\theta_0$ and depend on $s$ in the
%opposite case $\theta<\theta_0$.

Before presenting numerical results we would like to discuss the
asymptotic behaviour of the asymmetry $A$.
Let's discuss  the angular region defined by
(\ref{thetatu}). The Regge factors $(s/t)^{\omega}, (s/u)^{\omega}$
are equal to $\theta^{-2 \omega}$ in this region.
Eq.~(\ref{mpostu}) states that the asymptotics of all scattering
amplitudes $M_j^{(+)}$ is determined by the position of the
rightmost zero of the Parabolic cylinder function $D_{p_j}(z)$. The
location of the zeros of $D_{p_j}(z)$ depends on the value of $p_j$.  In
vicinity of the rightmost zero $z_0(p_j)$,
$D_{p_j}(z)$ can be represented as

\begin{equation}
\label{dzp}
D_{p_j}(z) = D'_{p_j}(z_0(p_j))(z - z_0(p_j))
\end{equation}

which, after substitution in Eq.~(\ref{mpostu}) leads to
the following expression for the high energy asymptotics of
$M_j^{(+)}$:

\begin{equation}
\label{asympt}
M^{(+)}_j \propto \theta^{- 2 \lambda_j z_0(p_j)} ~.
\end{equation}

When $p_j<0$, all zeros of $D_{p_j}(z)$ are in
the left half of the complex $z$ -plane and their
real parts are negative, so that
$M^{(+)}_j (\theta)$ oscillates and decreases when
$\theta \to
\theta_0$.
We remind that all $\lambda_j$ are positive.

When $0<p_j<1$, the rightmost zero becomes real but
still negative, so in this case $M^{(+)}_j (\theta)$   again decreases
when $\theta \to \theta_0$, though without oscillations.

Finally, when $p_j>1$, the rightmost zero is real
and positive, so that $M^{(+)}_j (\theta)$ increases when $\theta \to
\theta_0$.

On the other hand, in the angular region of small $\theta$

\begin{equation}
\label{smalltheta}
\theta_0 \geq \theta
\end{equation}

 in Eq.~(\ref{asympt})  $1/\theta^2$ must be replaced by
$s/\mu^2$.
Therefore, $M^{(+)}$ grows with $s$ in the angular region
(\ref{smalltheta}) when $p_j > 1$ and decreases when $p_j < 1$.

Basicaly, finding zeros of $D_p(z)$ functions is a rather tedious
procedure.
In order to simplify it, one can use the fact
that, at integer $p$,

\begin{equation}
\label{dh}
D_{p-1}(z)/D_p(z) = \sqrt{2} H_{p-1}(z/\sqrt{2})/H_p(z/\sqrt{2})
\end{equation}

where $H_p$ are the Hermite polynomials.
Now, finding the zeros of a polynomial is easier that for
$D_p(z)$.
As $p_j$ are expressed through $a_j, b_j, c_j$ (see Eq.~(\ref{fp})),
with $a_j, b_j$ and $c_j$ given in Table~1,  it is easy to see that
$M_{u,d}^F$ grow while $M_{u,d}^B$ fall when $\theta \to 0$.
Indeed, the small - $\theta$ asymptotics of $M^F_u$  is controlled by
the rightmost zeros of $D_{p_-}$ and $D_{p_+}$.
Eq.~(\ref{fp}) and Table~\ref{table2} give  $p_+ < p_-$ and

\begin{equation}
\label{puf}
p_u^F \equiv \max{[p_-,~p_+]} =
p_- = \frac{3g^2 - {g'}^2 Y_lY_q}{{g'}^2(Y_l + Y_q)^2} \approx 25~.
\end{equation}

The small - $\theta$ asymptotics of $M^F_d$  is controlled by the
rightmost zeros of $D_{p_0}$ and $D_{p_1}$. Eq.~(\ref{fp})  and
Table~\ref{table1} give $p_1 < p_0$ and

\begin{equation}
\label{pdf}
p_d^F \equiv \max{[p_0,~p_1]} =
p_0 = \frac{3g^2 - {g'}^2 Y_lY_q}{{g'}^2(Y_l + Y_q)^2} \approx 6~.
\end{equation}

\begin{figure}
\begin{center}
\begin{picture}(260,160)
\put(10,10){
\epsfbox{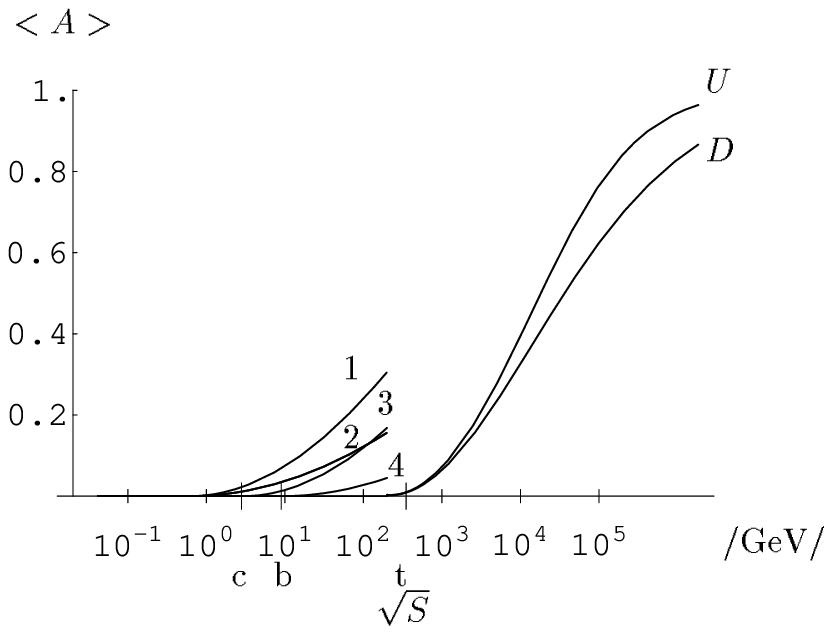}}
\end{picture}
\end{center}
\caption{
Charge asymmetry $A$ for $e^+e^-\to\bar{q}q$ for different quark flavors
in the ``collinear angular region''. Curves 1--4 were obtained
in pure QED (valid below threshold of EW bosons production):\\
\hspace*{5ex} 1 -- for $u$- quark with the infrared cut-off $\mu=300$~MeV;\\
\hspace*{5ex} 2 -- for $d$ and $s$ -quarks with the same cut-off;\\
\hspace*{5ex} 3 -- for $c$- quark with the infrared cut-off $\mu=1.4$~GeV;\\
\hspace*{5ex} 4 -- for $b$- quark with the infrared cut-off $\mu=4.5$~GeV.\\
Curves $U$ and $D$ correspond to any of $u$, $c$, $t$- quarks
and $d$, $s$, $b$- quarks respectively and were calculated in DLA for
EW theory with the infrared cut-off $\mu=100$~GeV.}
\label{fig3}
\end{figure}

Both $u$ - quarks in Eq.~(\ref{puf}) and $d$ - quarks in
Eq.~(\ref{pdf})
are left ones.
Amplitudes $M_u^B,~M_d^B$  are controlled by $D_p$ with $p_u^B \equiv
p_1$
and $p_d^B \equiv p_+$ respectively, and  both $p_u^B$ and $p_d^B$
are negative. Therefore, the asymmetry $A$, which is small
at relatively large $\theta$, grows up when $\theta$ decreases
down to $\theta_0$.  When the produced quarks are within the ``collinear''
angular region (\ref{smalltheta}), the asymmetry $A$ does not depend on
$\theta$ but turns to depend on $s$. Besides, the forward
amplitudes $M_{u,d}^F$ increase with $s$ in this angular region whereas
the backward amplitudes $M_{u,d}^B$ decrease. As a result, the asymmetry
$A$
is $s$ -dependent in the ``collinear'' angular region (\ref{smalltheta})
and increases up to 1 when $s\to\infty$.  Results of numerical
calculations of $A$ in ``collinear kinematics''  for different quark
flavours are presented in Fig.~\ref{fig3}. The curves below 100~GeV
were calculated in pure QED with the infrared cut-off $\mu=300$~MeV
for the case of light $u$, $d$ and $s$- quarks and with the
cut-off $\mu$ equal to quark mass for heavy quark flavours. Obviously, 
the value of the asymmetry is the same for the produced quarks and 
antiquarks (though they go in opposite directions). The energy
region in vicinity of $100$~GeV is close to the threshold of EW- boson
production and therefore the DL radiative corrections at such energies come 
from virtual photon exchanges whereas radiative corrections involving the 
weak bosons are non-logarithmic.  

Experimentally, quarks  manifest themselves through hadron jets.
Then to observe the effect of the asymmetry one could measure
asymmetry in production of a leading charged hadron ${\cal H}^{\pm}$
(a meson ${\cal M}^{\pm}$ or a barion ${\cal B}^{\pm}$) inside a narrow
cone along initial $e^-$ (or $e^+$) beam:

\begin{equation}
\label{23}
A({\cal H}) \equiv
\frac{d\sigma({\cal H}^-) - d\sigma({\cal H}^+)}{d\sigma({\cal H}^-) +
d\sigma({\cal H}^+)}~.
\end{equation}

\begin{figure}
\begin{center}
\begin{picture}(290,205)
\put(10,10){
\epsfbox{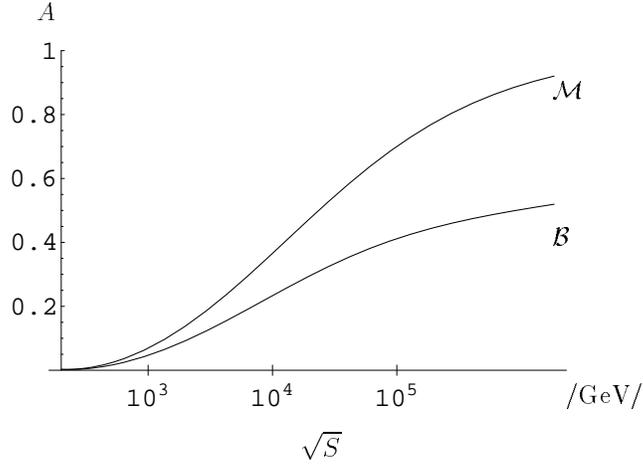}}
\end{picture}
\end{center}
\caption{
Estimation of charge asymmetry $A$ of leading charged hadrons in
$e^+e^-$ annihilation: The curve ${\cal M}$ is for the meson asymmetry
 and
the curve ${\cal B}$ is for the asymmetry of barions.}
\label{fig4}
\end{figure}

As ${\cal M}^-=\{\bar{U}D\}$ and ${\cal B}^-=\{\bar{U}\bar{U}\bar{D}\}$
where $U$ denotes any of $u$, $c$, $t$- quarks and $D$ stands for any of
$d$, $s$, $b$- quarks,
one should expect a more rapid rise with the annihilation energy $\sqrt{s}$
for $A({\cal M})$ than for $A({\cal B})$.
The reason is that at very high energies only $\bar{U}$ and $D$ are produced
in $e^-$- beam direction whereas production of $U$ and $\bar{D}$- quarks is
suppressed as the corresponding intercepts are negative.
Both $\bar{U}$ and $D$- quarks  can become constituent quarks of
a leading ${\cal M}^-$ and cannot be constituent quarks of a leading
${\cal M}^+$. Non-constituent quark contributions due to fragmentation
$\bar{U}\to{\cal M}^+$ and $D\to{\cal M}^+$
are subtracted in the numerator of Eq.~(\ref{23}) and can be suppressed
in the denominator by choosing the minimal momentum fraction $x_{min}$
in the definition of the leading hadron to be large enough.
For the case of
barion production a leading $\bar{U}$ can become a constituent quark
of ${\cal B}^-$, but now a leading $D$- quark can become a constituent
of ${\cal B}^+$ as well. So, a leading proton can also be observed in the
direction of $e^-$- beam.
At asymptotically high energies, the contribution of
$\bar{U}$- quark will dominate over contribution of $D$- quark and
$A({\cal B})$ will tend to unity, though slower than $A({\cal M})$~.
However the numerical estimations for high but finite energies shown in
Fig.~\ref{fig4}
demonstrate a sizable difference in the energy dependence between
$A({\cal B})$ and $A({\cal M})$.
It is worthwhile to note here that the phenomenological hadronization factors
for mesons,
$W(\bar{U}\to{\cal M}^-)=W(D\to{\cal M}^-)=W(U\to{\cal M}^+)=
W(\bar{D}\to{\cal M}^+)$~, are cancelled 
in  Eq.~(\ref{23}) for $A({\cal M})$~.
But the barion asymmetry $A({\cal B})$ involves the ratio

\begin{equation}
\label{rhf}
R = \frac{W(U\to{\cal B}^+)}{W(D\to {\cal B}^+)}~.
\end{equation}

The barion curve in Fig.~\ref{fig4} corresponds to $R=2.5$ which is in a good
agreement with $u$ and $d$- quarks fragmentation simulated with the 
JETSET\cite{s} for
$x_p > 0.3$ and $E_{quark}=200\div 2000$~GeV. A more detailed study of
hadron asymmetry with thorough account of nonleading and non-constituent
contributions will be done elsewhere.

The IREE~(\ref{eqfpos},\ref{eqfneg}) also describe $l\bar{l}$
-annihilation into another lepton-antilepton pair, with the lepton
being left, providing $Y_q$ is replaced by the lepton hypercharge.
Therefore, Eq.~(\ref{eqfpos}) can be used also for studying the
forward-backward asymmetry for $l\bar{l} \to l' \bar{l}'$
-annihilation.  Let us note that for such purely leptonic processes, the
forward (backward) kinematics is identical to $t$ $(u)$ - kinematics.

\section{Discussion}
Forward-backward asymmetry in $e^+ e^-$ annihilation into a
quark-antiquark pair has been considered in the double-logarithmic
approximation
at energies much higher than the masses of the weak bosons,
taking into account to all orders the exchange of virtual photons and $W,
Z$ -bosons.
In deriving our results
we have  not considered other channels than the annihilation into a
quark-antiquark pair.
Of course, other channels could be present, which could
change the cross section of the reaction, but will not contribute in DLA
to the forward-backward asymmetry. Therefore we expect, at very high
energies, that
accounting for other channels will not affect the asymmetry.
For example, besides EW effects, there are
perturbative QCD corrections to both the forward and backward annihilation.
In DLA, these corrections result into multiplying
$M_F$ and $M_B$ by the same Sudakov form factor

\begin{equation}
\label{sud}
S_{QCD}=\exp\left[-\,\frac{\alpha_s}{3\pi}
\ln^2\left(\frac{s}{\mu^2}\right)\right]
\end{equation}

which suppresses the soft gluon emission.
However, these Sudakov form factor effects cancel
when the gluon bremsstrahlung is taken
into account\cite{efl}. The same is true for the soft photon emission.

In addition to the soft emission,
there are
also important QED and QCD corrections accounting for
 hard photon and gluon bremsstrahlung\cite{el}.
Indeed, the impact of this ``harder''
emission should be considered in more detail.
Here we just notice that according to the results of
Ref.~\cite{el} accounting for such emission changes the arguments of $D_p$
in Eq.~(\ref{mpostu}) but does not change the position of
$z_0(p)$ and therefore it should not lead to big changes in our
formulae for the forward-backward asymmetry.
As stated earlier, the $l\bar{l}
\to l'\bar{l}'$ -annihilation in the kinematics $s\gg -u\sim M^2$
and with chiralities of the fermions corresponding to the particular
case $j=LR$  of Eqs.~(\ref{mpostu},\ref{right}) (corresponding to the EW
singlet amplitude) was considered in Ref.~\cite{flmm}.
The solution of the IREE for the positive signature amplitude of this
process obtained in that analysis coincides with our corresponding case
of Eqs.~(\ref{mpostu},\ref{right}).
Also, for annihilation at low cm energies, when
the asymmetry $A$ in QED is due to the DL
contributions from the multiphoton exchanges only, we agree with
Ref.\cite{gg}. It
is easy to obtain the
expression for the forward and backward QED amplitudes from
Eq.~(\ref{mpostu}), by putting  $g=0$, replacing ${g'}^2$ by $4 \pi
\alpha$ and substituting the hypercharges $Y_l$, $Y_q$ by the electric
charges $Q_l=-1$, $Q_q=e_q$.

\section{Acknowledgement}
We are grateful to V.A.~Khoze and 
M.G.~Ryskin  for discussion of effects of hadronization
to the barion asymmetry.
The work is supported by grants CERN/FIS/43652/2001, INTAS-97-30494,
RFBR 00-15-96610 and by EU QCDNET contract FMRX-CT98-0194.

\begin{table}[h]
\begin{center}
\begin{tabular}{|c|c|c|c|c|c||c|c||}
\hline
$F$ & $a_j$ & $b_j$ & $h_j$ & $c_j$ & $r_{j0} $ & $r_{j1} $ & $ p_j $
\\ \hline
$F_0$ & $\frac{3g^2\!+\!{g'}^2Y_lY_q}{4}$ &
${g'}^2\frac{(Y_l\!-\!Y_q)^2}{4}$ &
$\frac{3 g^2\!+\!{g'}^2Y_lY_q}{2}$ &
1 & ${g'}^2\frac{(Y_l\!+\!Y_q)^2}{4}$ & $3 g^2$ &
$\frac{3\!+\!Y_lY_q\tan^2\theta}{(Y_l\!-\!Y_q)^2\tan^2\theta}$\\
$F_1$ & $\frac{\!-\!g^2\!+\!{g'}^2Y_lY_q}{4}$ &
$2g^2\!+\!{g'}^2\frac{(Y_l\!-\!Y_q)^2}{4}$ &
$\frac{-g^2\!+\!{g'}^2Y_lY_q}{2}$ &
1 & $g^2$ & ${g'}^2\frac{(Y_l\!+\!Y_q)^2}{4}$ &
$\frac{\!-\!1\!+\!Y_lY_q\tan^2\theta}{8\!+\!(Y_l\!-\!Y_q)^2\tan^2\theta}$\\
$F_{RR}$ & $\frac{{g'}^2Y_lY_q}{4}$ &
${g'}^2\frac{(Y_l\!-\!Y_q)^2}{4}$ &
$\frac{{g'}^2Y_lY_q}{2}$ &
1 & ${g'}^2\frac{(Y_l\!+\!Y_q)^2}{4}$ & 0 &
$\frac{Y_lY_q}{(Y_l\!-\!Y_q)^2}$ \\
$F_{LR}$ &
$\frac{{g'}^2Y_lY_q}{4}$ &
$\frac{3g^2\!+\!{g'}^2(Y_l\!-\!Y_q)^2}{4}$ &
$\frac{{g'}^2Y_lY_q}{2}$ &
1 &
$\frac{3g^2\!+\!{g'}^2(Y_l\!+\!Y_q)^2}{4}$ & 0 &
$\frac{Y_lY_q\tan^2\theta}{3\!+\!(Y_l\!-\!Y_q)^2\tan^2\theta}$ \\
\hline \end{tabular}
\end{center}
\caption{\label{table1} The coefficients of IREE
Eqs.(\ref{eqfpos},\ref{eqfneg}) for $t$-kinematics.
The angle $\theta$ here
is the Weinberg angle.}
\end{table}

\begin{table}[h]
\begin{center}
\begin{tabular}{|c|c|c|c|c|c||c|c||}
\hline
$F$ & $a_j$ & $b_j$ & $h_j$ & $c_j$ & $ r_{j-}$ & $ r_{j+}$ & $p_j$
\\ \hline
$F_-$ & $\frac{\!-\!3g^2\!+\!{g'}^2Y_lY_q}{4}$ &
${g'}^2\frac{(Y_l\!+\!Y_q)^2}{4}$ &
$\frac{3 g^2\!-\!{g'}^2Y_lY_q}{2}$ &
-1 &
${g'}^2\frac{(Y_l\!-\!Y_q)^2}{4}$ & $3g^2$ &
$\frac{3\!-\!Y_lY_q\tan^2\theta}{(Y_l\!+\!Y_q)^2\tan^2\theta}$\\
$F_+$ & $\frac{g^2\!+\!{g'}^2Y_lY_q}{4}$ &
$2g^2\!+\!{g'}^2\frac{(Y_l\!+\!Y_q)^2}{4}$ &
$\frac{-g^2\!-\!{g'}^2Y_lY_q}{2}$ &
-1 &
$g^2$ & ${g'}^2\frac{(Y_l\!-\!Y_q)^2}{4}$ &
$-\frac{1\!+\!Y_lY_q\tan^2\theta}{8\!+\!(Y_l\!+\!Y_q)^2\tan^2\theta}$\\
$F_{RR}$ & $\frac{{g'}^2Y_lY_q}{4}$ &
${g'}^2\frac{(Y_l\!+\!Y_q)^2}{4}$ &
$\frac{-\!{g'}^2Y_lY_q}{2}$ &
-1 & ${g'}^2\frac{(Y_l\!-\!Y_q)^2}{4}$
& 0 & $-\frac{Y_lY_q}{(Y_l\!+\!Y_q)^2}$ \\
$F_{LR}$ & $\frac{{g'}^2Y_lY_q}{4}$ &
$\frac{3g^2\!+\!{g'}^2(Y_l\!+\!Y_q)^2}{4}$ &
$\frac{-\!{g'}^2Y_lY_q}{2}$ &
-1 &
$\frac{3g^2\!+\!{g'}^2(Y_l\!-\!Y_q)^2}{4}$ & 0 &
$-\frac{Y_lY_q\tan^2\theta}{3\!+\!(Y_l\!+\!Y_q)^2\tan^2\theta}$  \\
\hline \end{tabular}
\end{center}
\caption{\label{table2} The coefficients for IREE
Eqs.(\ref{eqfpos},\ref{eqfneg}) for $u$-kinematics.
The angle $\theta$ here is the Weinberg angle.}
\end{table}

%%%%%%%%%%%%%%%%%%%%%%%%%%%%%%%%
%%%%%%%%%%%%%%%%%%%%%%%%%%%%%%%%
%%%%%%%%%%%%%%%%%%%%%%%%%%%%%%%%
%%%%%%%%%%%%%%%%%%%%%%%%%%%%%%%%
%%%%%%%%%%%%%%%%%%%%%%%%%%%%%%%%

\end{document}